\documentclass[times,10pt,twocolumn]{article} 
\usepackage{latex8}
\usepackage{times}
 \usepackage[pdftex]{graphicx}
\usepackage{dblfloatfix}
\usepackage{framed}

\begin{document}

\title{Using the Agile Adoption Framework to Assess Agility and Guide Improvements}

\author{Lucas Gren\\
Chalmers and the University of Gothenburg \\ Gothenburg, Sweden 412--92
\\ lucas.gren@cse.gu.se\\
}

\maketitle

\thispagestyle{empty}

\section{Introduction}\label{ooo}

Agile software development has become an extremely popular way of managing projects, and there have also been studies showing the effectiveness of such approaches \cite{serrador}. Many different measurement models for agility have been proposed but according to \cite{ozcan} Sidky's Agile Adoption Framework \cite{sidkyphd} is the most complete to date. Many companies have implemented an agile approach but lack a way to evaluate the level of implementation, and need contextually specific tools \cite{xp20141}. However, getting some quantitative feedback from the teams could be a useful indication of agility that can be used to study (and improve) agile teams and also to compare them. This paper presents preliminary results from such a tool.

Sidky's \cite{sidkyphd} framework assesses the level of agility an organization is ready to implement and recommends what methods these should be. There are two main differences between Sidky's \cite{sidkyphd} tool and how we use it here. First, we measure the present level of agility and not agile potential. In order to measure agility level, the items were altered in time in our tool so they ask about the current situation. Second, it lets the group members fill out the survey and allows a statistical confidence interval to the result. One validation study has been conducted testing the tool in such a way \cite{grenjss}. The authors report that the feedback was considered useful by the team participating in the pretest, but that the measurement itself shows problems with validity. Despite that, and since the tool was considered useful by the team itself in the focus group, we believe our tool can be used to guide improvement efforts. Sidky \cite{sidkyphd}, also validated the content in the tool by letting practitioners evaluate the items and their connection to what they think agility is.

Sidky's framework is divided into ``agile levels'', ``principles'', ``practices and concepts'', and ``items or indicators''. 

To assess the results the Agile Adoption Framework includes 4 steps: First, calculating a weight for each item (the weight of 1 is divided by the number of items if all items are considered equally important), second, computing weighted intervals. These intervals are created by taking the answer of each item and calculating a pessimistic and optimistic result for each one, and the Likert scale is then divided into a percentage.

To clarify, this means that the practices ``Reflect and tune process'', ``Collaborative planning'', ``Collaborative teams'', ``Empowered and motivated teams'', ``Working standards\slash procedures'', ``Knowledge sharing tools'', ``Task volunteering'', and ``Customer commitment'' are assessed in the tool. The actual items (or indicators) are published in \cite{grenjss}. The tool consists of one survey for managers and one survey for developers. To assess an agile practice the analysis method proposed uses answers from both surveys. Some of the items are also used to assess more than one practice.

Below is the description of what the agile characteristics set out to determine. This list is needed to find what the different scores mean in the feedback table presented later.

\begin{table*}
\centering
\caption{The Result for Team A (the ``notes below table'' can be found in Section~\ref{ooo} of this paper).}
\label{fig:Company A}
\includegraphics[scale=0.35]{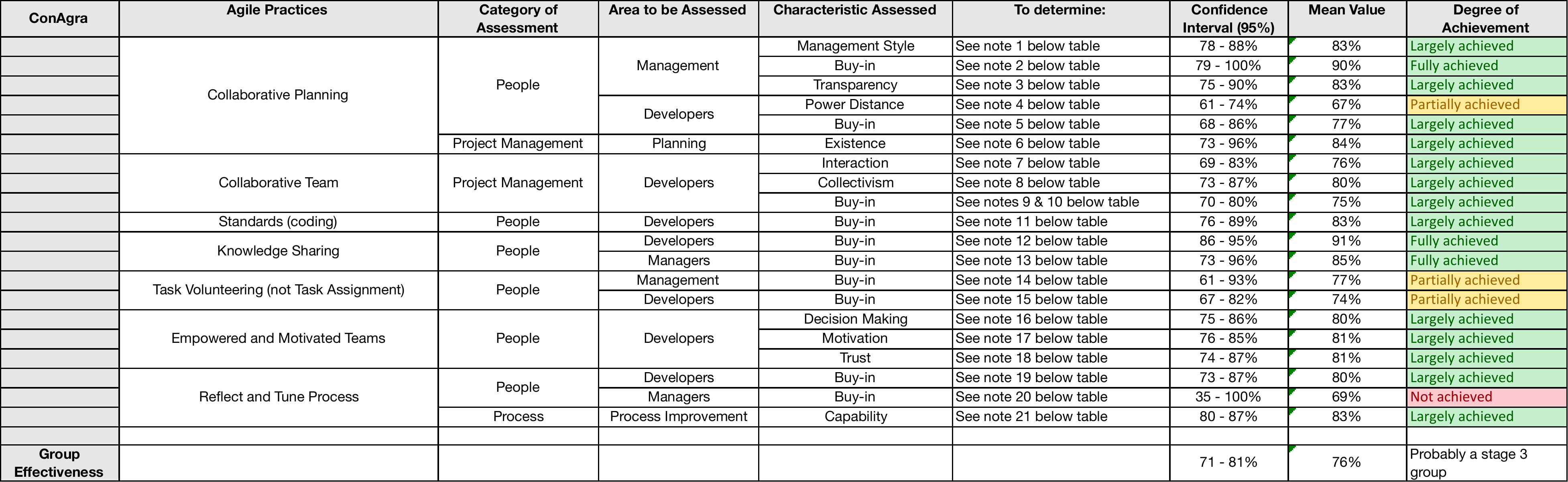}
\end{table*}


\begin{framed}
\scriptsize (1) Whether or not a collaborative or a command-control relation exists between managers and subordinates. The management style is an indication of whether or not management trusts the developers and vice versa. (2) Whether or not management is supportive of or resistive to having a collaborative environment. (3) Whether or not management can be open with customers and developers, i.e., no politics and secrets. (4) Whether or not people are intimidated\slash afraid to give honest feedback and participation in the presence of their managers. (5) Whether or not the developers are willing to plan in a collaborative environment. (6) Whether or not the organization does basic planning for its projects. (7) Whether or not any levels of interaction exist between people thus laying a foundation for more team work. (8) Whether or not people believe in group work and helping others or are just concerned about themselves. (9) Whether or not people are willing to work in teams. (10) Whether or not people recognize that their input is valuable in group work. (11) Whether or not the developers see the benefit and are willing to apply coding standards. (12) Whether or not developers believe in and can see the benefits of having project information communicated to the whole team. (13) Whether or not managers believe in and can see the benefits of having project information communicated to the whole team. (14) Whether or not management will be willing to buy into and can see benefits from employees volunteering for tasks instead of being assigned. (15) Whether or not developers are willing to see the benefits from volunteering for tasks. (16) Whether or not management empowers teams with decision making authority. (17) Whether or not people are treated in a way that motivates them. (18) Whether or not managers trust and believe in the technical team in order to truly empower them. (19) Whether or not developers are willing to commit to reflecting about and tuning the process after each iteration or release. (20) Whether or not management is willing to commit to reflecting about and tuning the process after each iteration or release. (21) Whether or not the organization can handle process change in the middle of the project.
\end{framed}


\section{Feedback to Companies}\label{sub:feedback}
So far, we have tested the tool on seven teams in three multinational companies consisting of two US-based and one European-based. Below is the result and recommendations that were given to each team based on the survey result. Only one result and feedback from the one team at Company A is shown in this paper as an example, but all teams were given feedback the same way.

\paragraph{Results and Recommendations for Company A --- One team}

Company A could focus on the following aspects to improve their agility (they are explained in more detail below): (1) Collaborative Planning with regard to the Power Distance between management and group members. (2) Task volunteering. (3) Management's buy-in of reflecting and tuning the process after each iteration or release.

Table~\ref{fig:Company A} shows the results for the team. 
Collaborative planning and its aspect of power distance got a lower score, which means the teams would benefit from having a flatter and less hierarchical organizational structure when planning projects. Having management plan and set up the projects without participation from team members, will result in less accurate and suboptimal goals. The team members are experts on how much work the team can do under a given period of time.

At Team A, ``task volunteering'' got a lower score from both mangers and developers. Task volunteering is also a key to success, since developers need to take full responsibility for their current chores and deliverables together with the team as a whole. Team members are often also better at knowing their capability and making time estimates on tasks.

The last aspect of Team A concerned manager's buy-in of ``Reflect and Tune Process''. This means that the management is not willing to reflect and tune the process after each iteration or release. Continuous improvement is a key to achieving as high performance as possible. Management of Team A would benefit from having a workshop on how agile principles are supposed to work in teams, especially how the process needs fine-tuning and adaptation to each team/or context. A practice that works well in one team might need to be adapted to a new context but with the reason behind it kept. I.e. the agile principle behind an implementation can be kept, but the actual implementation must be tuned to, and reflected on, in the new context. Hopefully this will guide them into accepting and seeing the benefits of reflecting and tuning the processes and task volunteering.

\section{Conclusions and Future Work}\label{sec:conclusions_and_future_work}
This paper has presented how the Agile Adoption Framework can be used to assess agility and pinpoint focus areas for companies that want to improve. Management found it useful in general to get data on possible focus points for improvement. Agility implies a set of principles that need to be followed in order to have the proposed responsiveness to change. We believe our tool could be useful as one step in the agile process assessment.

\nocite{ex1,ex2}
\bibliographystyle{latex8}
\bibliography{references}

\begin{thebibliography}{1}\setlength{\itemsep}{-1ex}\small

\bibitem{xp20141}
R.~M. Fontana, S.~Reinehr, and A.~Malucelli.
\newblock Maturing in agile: What is it about?
\newblock In G.~Cantone and M.~Marchesi, editors, {\em Agile Processes in
  Software Engineering and Extreme Programming (XP2014)}, volume 179 of {\em
  Lecture Notes in Business Information Processing}, pages 94--109. Springer,
  2014.

\bibitem{grenjss}
L.~Gren, R.~Torkar, and R.~Feldt.
\newblock The prospects of a quantitative measurement of agility: A validation
  study on an agile maturity model.
\newblock {\em Journal of Systems and Software}, 107:38--49, 2015.

\bibitem{ozcan}
O.~Ozcan-Top and O.~Demirors.
\newblock Assessment of agile maturity models: A multiple case study.
\newblock In T.~Woronowicz, T.~Rout, R.~O'Connor, and A.~Dorling, editors, {\em
  Software Process Improvement and Capability Determination}, volume 349 of
  {\em Communications in Computer and Information Science}, pages 130--141.
  Springer Berlin Heidelberg, 2013.

\bibitem{serrador}
P.~Serrador and J.~K. Pinto.
\newblock Does {Agile} work? {A} quantitative analysis of agile project
  success.
\newblock {\em International Journal of Project Management}, 33(5):1040--1051,
  July 2015.

\bibitem{sidkyphd}
A.~Sidky.
\newblock {\em A structured approach to adopting agile practices: The agile
  adoption framework}.
\newblock PhD thesis, Virginia Polytechnic Institute and State University,
  2007.

\end{thebibliography}

\end{document}